# Temporal Bell inequalities without noninvasive measurability

Ramon Lapiedra
Department of Astronomy and Astrophysics, University of Valencia, Spain

**Abstract**
   Some temporal Bell's inequalities are deduced under the assumptions of realism and perfect correlation. No locality condition is needed. The different measurements on the system refer to three external parameter values. In the case where the measured observable does not commute with the Hamiltonian, they also can refer to three different measurement times. When the system is a macroscopic one, the perfect correlation assumption substitutes advantageously the noninvasive measurability hypothesis. Some microscopic and macroscopic situations where these inequalities could be tested are considered.

### I. Introduction

   Aside the ordinary Bell inequalities [1,2] for entangled systems, there exist also what has been called temporal Bell inequalities [3] for a unique system. These inequalities have to do with temporal correlations instead of the space correlations present in the ordinary Bell inequalities. In Ref. [4], the authors consider a macroscopic system and assume that the state can be determined with an arbitrarily small perturbation on its subsequent dynamics. They call this assumption the "noninvasive measurability" (*a*). Furthermore, assuming realism (*b)*, these authors prove some temporal Bell inequalities for such a macroscopic system, where the measurement times, $t_i$, play the role of the polarizer settings in the ordinary Bell inequalities. It seems that assumption *a*) could be realized in actual experiments with macroscopic quantum systems [3,4]. Thus, assumptions *a*) and *b*) could be tested against the violation of those temporal inequalities, predicted by quantum mechanics, a violation that can be seen as a sort of entanglement in time.

   Here we consider an ensemble of systems *S*, prepared in some way at an initial time. *S* can be either macroscopic or not, and has a dichotomic magnitude, *M*, that is, a magnitude which only takes two values, say ±1. We will measure *M* for three different values, *a*, *b*, and *c*, of an external parameter. Furthermore, suppose we perform two immediately consecutive measurements for the same external parameter value. Then, we will assume that the two measurement results are *perfectly correlated*, i. e., if the first measurement value is +1, the second one is always +1, and similarly for the −1 value. A simple example is a *qubit* (i. e., a quantum system whose space of states is 2-dimensional) and the pairs of consecutive measurements done on it for three different external parameter values.

   In the present paper, we will prove some temporal Bell inequalities where the measurements times, $t_i$, of Ref. [4] will be replaced by the above parameter values, *a*, *b*, and *c*, present in the ordinary Bell inequalities. In Ref. [5], a similar problem is addressed. The authors claim to have proved the temporal analogous to the ordinary CHSH inequalities [2]. However, we believe that this pretended prove does not work, since their four types of measurements cannot be made at the same time. Hence, the authors should prove that they can couple the measurement results as they do in order to complete the prove, but this assumed coupling is not obvious at all.

## II. Proving the temporal Bell inequalities

Let us consider the above perfectly correlated system, $S$, with its dichotomic magnitude, $M$, measured randomly for the external parameter values $a$, $b$, and $c$. We want to prove a temporal Bell inequality for the results of these measurements, assuming realism and perfect correlation. In order to do so, we will adapt to our case a prove of an ordinary Bell inequality for a pair of entangled qubits in the singlet state, less restrictive that the original one of Bell [1]. We here adapt this prove in the form given by d'Espagnat [6], even if the prove was first given by Wigner [7]. The role played in this prove by the *perfect anti-correlation* of the singlet state will be played now by perfect correlation.

Let us be more precise on the kind of experiment we are going to consider. On each system, $S$, of the above ensemble, we perform two consecutive measurements of $M$ for two independent values, chosen randomly, of the three fixed external parameter values $a$, $b$, and $c$. Then, from the realism assumption, we will denote by $(a^\alpha b^\beta c^\gamma)$, where $\alpha, \beta, \gamma = \pm$, the reality such that, if we would do for the parameter $a$, or $b$, or $c$ the second of the above two measurements, we would obtain the result $\alpha 1$, or $\beta 1$, or $\gamma 1$, respectively. (Notice that, as d'Espagnat does in the Ref. [6], we assume the existence of a reality for all the possible results of all possible measurements, even if each measurement is done at one randomly chosen direction. In the case where system $S$ were entangled with a similar one, this would be the kind of reality assumed by Einstein, Podolsky, Rosen in their celebrated paper, i. e., the EPR elements of reality). We will denote by $N(a^\alpha b^\beta c^\gamma)$ the number of these realities, and we will define

$$N(a^+b^-) = N(a^+b^-c^+) + N(a^+b^-c^-), \qquad (2.1)$$
$$N(a^+c^-) = N(a^+b^+c^-) + N(a^+b^-c^-), \qquad (2.2)$$
$$N(b^-c^+) = N(a^+b^-c^+) + N(a^-b^-c^+). \qquad (2.3)$$

From this, we have:

$$N(a^+c^-) \leq N(a^+b^-c^-), \qquad N(b^-c^+) \leq N(a^+b^-c^+). \qquad (2.4)$$

Adding up these two inequalities and taking into account Eq. (2.1), it is straightforward to obtain the following inequalities

$$N(a^+b^-) \leq N(a^+c^-) + N(b^-c^+). \qquad (2.5)$$

Now, let us consider the number of *throwns*, $N[a^+,b^-]$, where $a^+$ is the result of the first measurement and $b^-$ the result from the second one. Obviously, these *throwns* can only come from the realities $(b^-)$. Furthermore, from the perfect correlation assumption, they can only come from the more specific realities $(a^+b^-)$. (The notation $(b^-)$ and $(a^+b^-)$ should be obvious). Then, given a reality such as $(a^+b^-)$, which is the probability of obtaining a *thrown* as $[a^+,b^-]$? Since the choice of one of the three parameters, $a$, $b$, $c$, is a random choice, this probability is just *1/9*. This means that we can write

$$N(a^+b^-) = 9N[a^+,b^-], \qquad (2.6)$$

and similarly for $N(a^+c^-)$ and $N(b^-c^+)$. Thus, taking into account Eq. (2.5), we obtain the temporal Bell inequality:

$$N[a^+,b^-] \leq N[a^+,c^-] + N[b^-,c^+], \qquad (2.7)$$

for the observable quantities $N[a^+,b^-]$, $N[a^+,c^-]$, and $N[b^-,c^+]$. Obviously, one can obtain similar different inequalities interchanging here signs and parameter values in a suitable way.

If we prefer to speak in terms of probabilities corresponding to the numbers in ineq. (2.7), we can write this inequality as

$$P(a^+,b^-) \leq P(a^+,c^-) + P(b^-,c^+), \qquad (2.8)$$

and similarly
$$P(a^-,b^+) \leq P(a^-,c^+) + P(b^+,c^-). \tag{2.9}$$
But, for the expected value,
$$E(a,b) = P(a^+,b^+) + P(a^-,b^-) - P(a^+,b^-) - P(a^-,b^+), \tag{2.10}$$
we have trivially
$$E(a,b) = 1 - 2[P(a^+,b^-) + P(a^-,b^+)]. \tag{2.11}$$
Thus, taking into account ineqs. (2.8) and (2.9), we obtain finally:
$$E(a,c) + E(b,c) - E(a,b) \leq 1, \tag{2.12}$$
which is an interesting version of our temporal Bell inequalities, as we will see in the next Section.

It is easy to see that one also arrives to ineqs. (2.8), (2.9), and (2.12) when the three measurements of the magnitude *M*, for three different parameter values, *a*, *b*, and *c*, are substituted by three measurements of *M*, each one of them done at a time randomly chosen among three previously fixed times. (This is what is done, incidentally, by Legget and Garg in their seminal paper [4], where they introduce some macroscopic temporal Bell inequalities). This equivalence between parameter values and time values is true provided that magnitude *M* does not commute with the Hamiltonian of the system, otherwise the three different times would be equivalent to the same parameter value.

Ineq. (2.12) can be compared with the original Bell inequality for a pair of correlated spin ½ particles in the singlet state:
$$/E(a,b) - E(a,c)/ - E(b,c) \leq 1. \tag{2.13}$$
To prove this inequality one assumes a kind of light realism, i.e., behind each measurement which is actually done there exists the corresponding reality, instead of the more restrictive realism required to prove ineq. (2.12). Nevertheless, ineq. (2.12) is less restrictive than ineq. (2.13), since if the latter is satisfied than the ineq. obtained by suppressing in (2.13) the absolute value symbol will also be satisfied, but not the reverse case. The reason why ineq. (2.13) becomes more restrictive than ineq. (2.12) is that, in proving (13) there is an assumption which was not needed to prove (2.12), namely, that the result of the second of the two measurements of any *thrown* cannot be influenced by the result of the corresponding first measurement. However, this cannot be assumed to prove (2.12) since our *thrown* consist of two consecutive measurements on the same system.

At first sight, one could think that ineq. (2.12) has no interest since, if it were experimentally violated, this could always be explained by some transmission of information between the two consecutive measurements of the *thrown*. But this is not true since, as we have seen, ineq. (2.12) can be deduced from the assumptions of realism and perfect correlation, without any further assumptions. Therefore, we can transmit all kind of information we want between both measurements, or even assume any supplementary correlations between them, but if realism and perfect correlation are preserved, as we assume, ineq. (2.12) must remain true. Thus, if this inequality were experimentally violated, this would imply that the hypothesis of realism must be descarded, irrespective of the information that could have been transmitted between the two consecutive measurements, as well as of any hypothetical supplementary correlations between them.

These claims can also be sustained if the Wigner inequalities [7], for two entangled qubits at the singlet state, were experimentally violated, which is a remarkable fact.

### III. Quantum violation of the temporal Bell inequalities

Let us assume that our system *S* is a qubit. Its general state, $|\psi\rangle$, can be written:
$$|\psi\rangle = s\,|e+\rangle + (1-s^2)^{1/2}\,e^{i\varphi}\,|e-\rangle, \qquad (3.1)$$
where $|e+\rangle$ and $|e-\rangle$ are the eigenstates of eigenvalues $\pm 1$, respectively, for a given "direction" *e*. Then, since for any "direction" *x* the corresponding eigenstates, $|x+\rangle$ and $|x-\rangle$, are orthogonal unit vectors in a 2-dimensional Hilbert space, it is straightforward to show that it always exists an angle $\alpha_x$ such that
$$|x+\rangle = [(1+\cos\alpha_x)/2]^{1/2}|e+\rangle + [(1-\cos\alpha_x)/2]^{1/2}|e-\rangle, \qquad (3.2)$$
$$|x-\rangle = [(1-\cos\alpha_x)/2]^{1/2}|e+\rangle - [(1+\cos\alpha_x)/2]^{1/2}|e-\rangle. \qquad (3.3)$$
This means that, as it is well-known, *x* and *e* can always be interpreted as two unit 3-vectors in $\mathbf{R}^3$, **x** and **e**, respectively, which appear at Eqs. (3.2) and (3.3) only through their 3-scalar product $\mathbf{x}\cdot\mathbf{e} = \cos\alpha_x$.

Hence, when measuring the above dichotomic magnitude *M* for the three external parametervalues, *a*, *b*, and *c*, we can always say that these measurements have been done for the corresponding unit 3-vectors, **a**, **b**, and **c**.

Let us consider the probability, $P(\mathbf{a}^\pm,\mathbf{b}^\pm)$, of obtaining $\pm 1$ for the two consecutive measurements of the *thrown* where the chances have selected, respectively, the unit 3-vectors **a** and **b**. After some basic algebra, we find
$$P(\mathbf{a}^+,\mathbf{b}^+) = s^2(1+\mathbf{a}\cdot\mathbf{b})/2, \quad P(\mathbf{a}^-,\mathbf{b}^-) = (1-s^2)(1+\mathbf{a}\cdot\mathbf{b})/2. \qquad (3.4)$$
But, similarly to Eq. (2.11), we can write:
$$E(\mathbf{a},\mathbf{b}) = 2[P(\mathbf{a}^+,\mathbf{b}^+)+P(\mathbf{a}^-,\mathbf{b}^-)] - 1. \qquad (3.5)$$
Thus, we obtain:
$$E(\mathbf{a},\mathbf{b}) = \mathbf{a}\cdot\mathbf{b}, \qquad (3.6)$$
that differs in one sign from the similar result for the expected value in the case of an entangled pair of qubits in the singlet state. As it was noticed in Ref. [5], this result has the remarkable property of being independent of the initial state of the particle, that is, in Eq. (19), $E(\mathbf{a},\mathbf{b})$ does not depend on *s* or $\varphi$ appearing in Eq. (3.1), while $P(\mathbf{a}^\pm,\mathbf{b}^\pm)$ does. This is the reason why in the above Section it was stated that inequality (2.12) is a useful version of our temporal Bell inequalities, since this version is independent of the initial state of the system *S*. (Obviously, for $E(\mathbf{a},\mathbf{c})$ and $E(\mathbf{c},\mathbf{b})$, we have similar equations to eq. (3.6)).

Keeping in mind Eq. (3.6) and the similar ones, the Bell inequality (2.11) becomes
$$(\mathbf{a}+\mathbf{b})\cdot\mathbf{c} - \mathbf{a}\cdot\mathbf{b} \leq 1, \qquad (3.7)$$
which is maximally violated by any two orthogonal unit 3-vectors **a** and **b**, if the unit 3-vector **c** is collinear to $\mathbf{a}+\mathbf{b}$. In all these cases the left hand side of inequality (3.5) takes the value $\sqrt{2}$.

### IV. Microscopic examples

Once we have seen that the temporal Bell inequalities (2.12) can be violated by quantum mechanics, we turn to the question of how this violation could be experimentally produced. Here, the problem is that we need to perform two successive measurements on the same unique system, and not merely on two distinct parts of the same system, as in the ordinary space entangled Bell inequalities. Then, we must guarantee that the first of these two measurements always be a measurement of first class, i. e., a preparation, in order to preserve the existence of the system and be able of doing the second measurement.

These conditions can be easily fulfilled in the case were the measured system is a ½ spin particle, whose spin is successively measured along different directions. On this point one can follow the strategy due to Tesche, as it is quoted in Ref. [3]. But this conditions can also be fulfilled in other less obvious cases, as for example in the case of the *B*-meson system considered at Ref. [8]. Here, the authors claim to have observed a violation of a Bell's inequality using particle-antiparticle correlations in semi-leptonic *B*-mesons decay. At the initial time, they produce the entangled state

$$\Psi = \frac{(B_l^0 \overline{B}_r^0 - \overline{B}_l^0 B_r^0)}{\sqrt{2}} \qquad (4.1)$$

of the particle-antiparticle pairs $B^0 \overline{B}^0$, where *l* and *r* mean left hand and right hand, respectively. This initial state is formally identical to the singlet state for two entangled ½ spin particles; yet, because of decay rates and the flavor $B^0 - \overline{B}^0$ oscillations in time, it evolves in time in free space.

Then, we will consider this initial state, which means that we will do the measurements just after the corresponding initial time. Very broadly speaking, to have a feasible experiment where the temporal Bell inequalities can be violated, we must do measurements on one of the two sides of the entangled state (4.1), for example, the left hand *B* mesons of this state. These measurements must be done in three different directions in the inner space spanned by the two orthogonal states $B^0$ and $\overline{B}^0$. To do so we will observe three different decay modes through its corresponding decay products. Notice that, differently to the case of the neutral kaons considered at Ref. [9], we can, for B mesons, do the above three different kinds of measurements because we have many different decay channels we can use. Obviously, it could happen that among these different "directions" we could not find three of them such that the corresponding expected values violate Bell's inequalities.

Now, as we have pointed out before, we need that the first of the two consecutive measurements of the same couple of measurements always be a first class measurement, i. e., a preparation. But, obviously, the above decays are by no means a preparation. To circumvent this difficulty we will do these first measurements by observing on the right hand the anti-correlated state of the eigenstate we want observe on the left.

In the event one could overcome the obvious practical difficulties which might appear in its actual performance, which could be the interest of doing an experiment as the one we have schematically described? To begin with, it has been correctly claimed in Ref. [10] that the experiment analyzed by Ref. [8] is not suitable for Bell tests since one cannot guarantee that the outcomes of the measurements on one side are not affected by the experimental setting chosen to measure on the other side. But, as we have explained above, the prove of the temporal Bell's inequalities we are considering here do not rely on the locality assumption. Instead, the prove only depends on the realism assumption and the observed perfect correlation. Thus, if these temporal inequalities were experimentally violated, it would be just realism which would be disregarded (without further concerns about locality conditions, contrary to the case treated in Ref. [8]). Recall, nevertheless, that the kind of reality which would be disregarded in this way is the EPR "elements of reality".

**V. Macroscopic examples**

As we have mentioned at the Introduction, under the assumption of both, realism and perfect correlation, the temporal Bell's inequalities (2.12) are also valid for any macroscopic system, *S*, with a dichotomic random magnitude, *M*. Here, the perfect correlation assumption substitutes the "noninvasive measurability" [4].

Obviously, the first difficulty in trying to apply these inequalities to a macroscopic (or mesoscopic) system, *S*, is to assure that the system fulfills the perfect correlation assumption. According to Eq. (3.6) and to the other similar equations, this assumption, in terms of the observed $M$ expected values, $E(x_n,x_m)$, $x_1 = a$, $x_2 = b$, $x_3 = c$, $n,m=1,2,3$, reads

$$E(x_n,x_n) = 1. \qquad (5.1)$$

One could now ask why the corresponding inequalities (2.12) could be violated for some macroscopic system. One reason could be that the macroscopic system $S$ encloses some quantum system (e. g., a mesoscopic quantum system, similar to that experimentally reported by [11] recently) whose dichotomic outcomes become macroscopically amplified by the macroscopic (or mesoscopic) system, $S$, in the same way a macroscopic measurement device does [12]. In this case, by Eqs. like (3.6), it follows that the observed expected values, $E(x_n,x_m)$, must fulfill conditions (5.1) and also the symmetric conditions:

$$E(x_n,x_m) - E(x_m,x_n) = 0. \qquad (5.2)$$

Next, let us suppose that conditions (5.1) and (5.2) are only nearly satisfied by the observed expected values, $E(x_n,x_m)$. That is, instead of these exact conditions we will have:

$$E(x_n,x_n) = 1 + O_n(\delta), \quad E(x_n,x_m) - E(x_m,x_n) = O_{nm}(\delta), \qquad (5.3)$$

where $\delta$ stands for some suitable small quantity, $\delta < 1$, and $O_n(\delta)$ and $O_{nm}(\delta)$ are of order $\delta$. But, according to its definition (see Eqs. (2.10) and (2.11)), the absolute value of the expected values $E(x_n,x_m)$ is less than 1. Therefore, Eq. (3.6) and the similar ones will become

$$E(x_n,x_m) = \boldsymbol{x_n}.\boldsymbol{x_m} + O_{nm}(\delta), \qquad (5.4)$$

where $\boldsymbol{x_n}.\boldsymbol{x_m}$ denotes the symmetric part of $E(x_n,x_m)$ with respect to the $n,m$ indices. Following this notation, this means that ineq. (2.12) becomes

$$(\boldsymbol{a}+\boldsymbol{b}).\boldsymbol{c} - \boldsymbol{a}.\boldsymbol{b}) + O_{ac}(\delta) + O_{bc}(\delta) - O_{ab}(\delta) \leq 1. \qquad (5.5)$$

Sice $O_{nm}(\delta)$ is anti-symmetric in indices $n, m$, by interchanging indices $a$ and $b$ at ineq. (5.5), and adding the inequality so obtained to (5.5), we obtain

$$(\boldsymbol{a}+\boldsymbol{b}).\boldsymbol{c} - \boldsymbol{a}.\boldsymbol{b}) + 2\,O(\delta) \leq 1, \qquad (5.6)$$

where $O(\delta)$ is generically of order $\delta$.

On the other side, it has been reported at the end of Section III that the maximum value of $(\boldsymbol{a}+\boldsymbol{b}).\boldsymbol{c} - \boldsymbol{a}.\boldsymbol{b}$ is $\sqrt{2}$. Then, in the ideal case in which there were no any other measurement errors than those tied to $\delta$, the maximum allowed value of $\delta$ to detect some possible violation of ineq. (5.6) (i. e., some violation of the temporal Bell's inequality (2.12)) is 0.2, roughly speaking.

Now, where could we find a mesoscopic dichotomic system for which the near perfect correlation condition,

$$E(x_n,x_n) = 1 + O_n(\delta), \qquad (5.7)$$

and the other conditions at Eq. (5.3), were satisfied with $\delta \leq 0.2$? It seems, perhaps, that it could be found in the realm of neurons, either in the case of a simple neuron, or in the case of a brain slice made of a few thousands of neurons, or even in the case of the encephalogram response of a living brain to some external and well defined excitations [13]. Exploring this kind of possibilities would have the following particular interest: people have speculated *ad nauseam* about the possibility that behind human consciousness, and perhaps behind life itself, quantum mechanics could be at work in some way. If it were possible to find an example of such dichotomic system in this realm of the neurons, in order to test the above temporal Bell's inequalities, the example should be welcome, irrespective of the violation of the inequalities, since this could be a way of experimentally exploring the possibility of a quantum origin for some life manifestations. The subject deserves certainly more attention and this is challenging work deferred for the future.

## VI.-Conclusions

In the present paper, we have proved some temporal Bell inequalities under the assumptions of EPR realism and perfect correlation, for any kind of physical system with a dichotomic magnitude, i. e., a magnitude which only can take two values. The measurement outcomes are the response of the system to some different external parameter values, as in the standard Bell's inequalities, or to different measurement times, as in the seminal paper of Legget and Garg [4], where the authors consider the assumption of "noninvasive measurability" in the context of macroscopic systems. In the present paper, when the physical system is a macroscopic one, the perfect correlation assumption substitutes advantageously the above non-invasive measurability.

The expected values which appear in the new temporal Bell's inequalities have to do with pairs of consecutive measurements made on the system as such, and not on different parts of an entangled system, as in the ordinary Bell's inequalities, which means that one has to deal with a sort of temporal entanglement. Notice that no locality assumption is needed to prove the present temporal Bell's inequalities. Furthermore, when trying to apply it to a given macroscopic system, we need only be sure that the perfect correlation assumption holds or is nearly satisfied, while we need not be concerned with any kind of information which could be propagated between two successive measurements.

The experimental violation of these temporal Bell inequalities would mean the failure of the EPR "elements of reality", that is, the failure of a kind of reality so exigent as to put something preexisting behind the feasible measurement results, and not the more modest sort of reality which would be behind the measurements actually done.

In particular, the present temporal Bell's inequalities apply in particular to the case of a ½ spin particle. In this case the expected values refer to the pairs of successive measurements of spin on three fixed directions, when the direction of each measurement is chosen randomly among the three above directions. It is easily proved that quantum mechanics violates these inequalities for some directions.

Since the experimental violation of these inequalities is interesting by itself, we have discussed some other scenarios where such a possible violation could be tested. More precisely, in the microphysics domain, we have proposed to consider the case of an entangled state of particle-antiparticle *B* mesons, which are nowadays routinely produced at the laboratory. In the macroscopic domain we have conjectured on the possibility of some good examples in the realm of neurons, where our temporal Bell's inequalities could potentially shed some light on the always revisited question of life and quantum mechanics, from an experimental point of view.

**Acknowledgements**

This work has been supported by the Spanish MCyT (Project AYA 2003-08739-C02 partially founded with FEDER) and also by the Generalitat Valenciana (grupo 03/170).

I would like to thank to Prof. Francisco Botella, from the Department of Theoretical Physics, of the University of Valencia, for some useful suggestions about decaying B mesons and Bell's inequalities.